# Hyperbolic Resonances of Metasurface Cavities

D. Keene[1], M. Durach[1,*]

[1] *Georgia Southern University, Statesboro, GA 30458*

*[mdurach@georgiasouthern.edu](mailto:mdurach@georgiasouthern.edu)*

**We propose a new class of optical resonator structures featuring one or two metasurface reflectors or *metacavities* and predict that such resonators support novel *hyperbolic resonances*. As an example of such resonances we introduce hyperbolic Tamm plasmons (HTPs) and hyperbolic Fabry-Perot resonances (HFPs). The hyperbolic optical modes feature low-loss incident power re-distribution over TM and TE polarization output channels, clover-leaf anisotropic dispersion, and other unique properties which are tunable and are useful for multiple applications.**

Planar optical cavities and resonators are an integral part of optoelectronic components, providing opportunities for control over field distribution, enhancement, and localization on the nano- and microscale. Novel types of optical resonances have always attracted attention of the photonics community. As an example, the recent theoretical proposal[1] and experimental realization[2] of Tamm plasmon (TP) polaritons has been a source for a magnitude of studies and novel devices, ranging from compact lasers[3] to exiton-polariton BEC vortex guns[4]. Simple TP resonances are formed on an interface between a metal and a DBR and are optical surface waves, but there is a range of coupled hybrid resonances of TP nature[5-8]. In particular, it has been recently demonstrated that TPs are related to Fabry-Perot (FP) resonances and continuously transform into them when a metal-DBR-metal (M-DBR-M) structure is modified into a metal-insulator-metal (MIM) structure via reduction of dielectric contrast in the DBR[8]. An important avenue in the research on TPs is consideration of TP formation on interfaces between DBRs and localized metal nanostructures[9].

Interesting examples of recently proposed optical cavity resonances include ones in which the cavity contains a hyperbolic metamaterial core[10] or a cavity formed between a metasurface and a reflector[11, 12]. Metamaterials, formed by optical systems which are structured on the subwavelength scale, have been one of the central topics in photonics. One of the common examples of metamaterials are one-dimensional metal-dielectric arrays, which exhibit a wide range of phenomena, including various optical phases[13]. Generally modeled as uniaxial anisotropic media, depending on signs of real parts of dielectric permittivities, the uniaxial metamaterials can behave as anisotropic dielectrics, hyperbolic materials, or anisotropic metals. With this range of optical properties comes a large variety of optical resonant responses supported by these metamaterial structures[14].

In this paper we propose a new class of cavities formed by a thin uniaxial metamaterial film (a uniaxial metasurface) and a conventional reflecting surface (Type I) or by two uniaxial metasurfaces (Type II) and show that such *metasurface cavities* (or *metacavities*) support novel resonances, which feature unique polarization properties, field distributions and anisotropic dispersion.

**Generic Optical Properties of Structures Containing Uniaxial Metasurfaces**

**Uniaxial Metasurfaces.** First, let us consider the general properties of structures containing uniaxial metasurfaces. In this paper we model such metasurfaces as thin films with thickness $h$ of an anisotropic uniaxial medium with a diagonal dielectric tensor, whose component along the optic axis is $\varepsilon_\parallel$ and is $\varepsilon_\perp$ in the perpendicular directions. Representing each metasurface this way with the optic axis directed in the plane of the metasurface serves as an approximation for a metal-dielectric grating, whose stratification axis coincides with the optic axis of the metasurface. The dielectric permittivity of the metasurface is defined in terms of the grating as $\varepsilon_\perp = \varepsilon_m f + \varepsilon_d (1-f)$ and $\varepsilon_\parallel^{-1} = \varepsilon_m^{-1} f + \varepsilon_d^{-1}(1-f)$, where $\varepsilon_m$ is the permittivity of the metal and $\varepsilon_d$ is the permittivity of the dielectric, while $f$ is the volumetric metal fraction[15].

**Characteristic Matrix Approach.** We use two coordinate systems interchangeably: one with the x-axis going along the optic axis and another with an $x'$-axis lying at the intersection of the incidence plane and the $xy$-plane. We denote the angle between the x- and $x'$-axes as $\phi$. The tangential components of the electric and magnetic field vectors are continuous throughout the structure and at different locations in the structure are related through matrix equations of the form

$$\begin{pmatrix} H_{y'}(z_0) \\ E_{x'}(z_0) \\ E_{y'}(z_0) \\ H_{x'}(z_0) \end{pmatrix} = \widehat{M} \begin{pmatrix} H_{y'}(z) \\ E_{x'}(z) \\ E_{y'}(z) \\ H_{x'}(z) \end{pmatrix}, \qquad (1)$$

where field components are taken in the $x'y'z$ coordinates.

For concreteness we consider transverse-magnetically (TM) polarized excitation. Due to the anisotropic response of the metasurface pure transverse-magnetically polarized incidence generally produces transverse-electrically (TE) polarized fields as well. The first two components of the vectors in Eq. (1) correspond to the TM fields, while the second pair is the TE fields. The equation which describes the reflection and transmission properties of the structure in this case can be written as

$$\begin{pmatrix} 1 + r_p \\ p_{p0}(1 - r_p) \\ r_s \\ -p_{s0} r_s \end{pmatrix} = \widehat{M} \begin{pmatrix} t_p \\ p_{ps} t_p \\ t_s \\ p_{ss} t_s \end{pmatrix}, \tag{2}$$

where $r_p, t_p$ and $r_s, t_s$ are the reflection and transmission coefficients in TM and TE polarizations correspondingly, while $p_p = k_z/k_0 \varepsilon$ and $p_s = -k_z/k_0$ with extra indices 0 and $s$ denoting the incidence medium and the substrate. The characteristic matrix of the structure $\widehat{M}$ is a $4 \times 4$ matrix[16]. To construct $\widehat{M}$ we use a modified characteristic matrix approach which takes into account the anisotropic nature of the structure and coupling between the TM and TE polarizations. The matrices for isotropic media are block matrices composed out of $2 \times 2$ characteristic matrices positioned along the main diagonal, with blocks of zeroes off the main diagonal, representing the absence of TM/TE coupling in isotropic materials. The structure of characteristic matrices of the metasurfaces is more complex and must contain these off-diagonal elements to account for the coupling of TM and TE fields in the metasurface layer. We provide the detailed description of those matrices in the Supplementary Information.

**Polarization Rotator Equations.** To understand the optical response of the structures consider first an excitation with incidence plane oriented along the optic axis ($\phi = 0°$). In this case the electric field is along the optic axis. It is determined by $\varepsilon_\parallel$, is independent of $\varepsilon_\perp$, and all the fields in the structure are TM polarized. Similarly, in response to a normal TM incidence with $\phi = 90°$ the electric field is TM polarized and is determined by $\varepsilon_\perp$, not $\varepsilon_\parallel$. As we show in the Supplementary Information, both the TM and TE reflection and transmission coefficients at arbitrary angle $\phi$ can be expressed through the reflection and transmission coefficients of the TM polarized fields at $\phi = 0°$ and $\phi = 90°$ as

$$r_p^\phi = r_p^{0°} \cos^2 \phi + r_p^{90°} \sin^2 \phi, \qquad t_p^\phi = t_p^{0°} \cos^2 \phi + t_p^{90°} \sin^2 \phi,$$

$$r_s^\phi = \left(r_p^{0°} - r_p^{90°}\right) \sin \phi \cos \phi, \qquad t_s^\phi = \left(t_p^{90°} - t_p^{0°}\right) \sin \phi \cos \phi. \tag{3}$$

These results, which we call Polarization Rotator Equations, imply that if the reflection and transmission coefficients are different for $\phi = 0°$ and $\phi = 90°$, then for the case of the incidence plane being at an intermediate angle, there should be an admixture of TE polarized response to purely TM polarized incidence, i.e. partial 90° polarization rotation. This gives way to a new class of resonances which we predict in structures containing uniaxial metasurfaces.

**Metacavity Resonances**

**Type I Metacavities.** Consider a layer of homogeneous material between a conventional reflective structure with reflection coefficients $r_p^R$ and $r_s^R$ in TM and TE polarizations and a uniaxial metasurface with reflection coefficients $r_{pp}^L$ and $r_{sp}^L$ in TM polarization in response to TM and TE incidence correspondingly and $r_{ps}^L$ and $r_{ss}^L$ in TE polarization (see Fig. 1(a)). In this case the resonance condition can be found from the following matrix equation

$$\begin{pmatrix} A_p \\ A_p\, r_{pp}^L + A_s\, r_{sp}^L \\ A_s \\ A_p\, r_{ps}^L + A_s\, r_{ss}^L \end{pmatrix} = \begin{pmatrix} e^{i\varphi} & 0 & 0 & 0 \\ 0 & e^{-i\varphi} & 0 & 0 \\ 0 & 0 & e^{i\varphi} & 0 \\ 0 & 0 & 0 & e^{-i\varphi} \end{pmatrix} \begin{pmatrix} B_p\, r_p^R \\ B_p \\ B_s\, r_s^R \\ B_s \end{pmatrix}. \tag{4}$$

In Eq. (4) the first and third rows are the amplitudes of the wave propagating to the right in TM and TE polarizations correspondingly, while the second and the forth are amplitudes of the left propagating waves. The vector on the left-hand side is composed of the values of those amplitudes on the left boundary of the metacavity (metasurface), while the vector on the right is the amplitudes at the right side of the metacavity (conventional reflector). They are connected by a transfer matrix, where $\varphi$ is the propagation phase inside of the cavity. Eq. (4) has nontrivial solutions only if

$$\left(r_{pp}^L + r_{ps}^L \frac{r_s^R r_{sp}^L e^{2i\varphi}}{1 - r_{ss}^L r_s^R e^{2i\varphi}}\right) r_p^R e^{2i\varphi} - 1 = \left(r_{ss}^L + r_{sp}^L \frac{r_p^R r_{ps}^L e^{2i\varphi}}{1 - r_{pp}^L r_p^R e^{2i\varphi}}\right) r_s^R e^{2i\varphi} - 1 = 0, \tag{5}$$

This expression describes the resonances of a metacavity which includes a single uniaxial metasurface (Type I). It is different from the resonances of conventional cavities, described by $r^L r^R e^{2i\varphi} = 1$. These novel resonances involve co-propagation and multiple coupled reflections of TM and TE polarized waves as shown in Fig 1(a).

**Hyperbolic Tamm Plasmons.** To give an example of such a resonance, we describe in detail the hyperbolic Tamm plasmons (HTPs), the existence of which we have recently predicted[17]. A schematic of the structure which supports HTPs is shown in Fig. 1(b). It is composed of two metasurfaces and a DBR. The metasurfaces and the DBR are separated by spacers with index of refraction $n_h$ and thickness $d_v$. The $(LH)^N$ DBR array has Bragg frequency $\omega_B$, with layers of a variable index of refraction $n_l$ and a set index of refraction $n_h = 3.6$, such that $n_h > n_l$. This arrangement corresponds to GaAs/GaAlAs DBR.

The thicknesses of the corresponding layers are $d_l = \pi c/(2n_l \omega_B)$ and $d_h = \pi c/(2n_h \omega_B)$. The metasurfaces go through a variety of optical phases as the metal fraction $f$ of the metasurfaces is changed, which is determined by the relationship between $\varepsilon_\parallel$ and $\varepsilon_\perp$. Decreasing the metal fraction $f$ from 1 turns the metasurface first into an anisotropic metal, since both $\text{Re}\,\varepsilon_\parallel$ and $\text{Re}\,\varepsilon_\perp$ are negative. At intermediate values of $f$ the metasurface behaves as a hyperbolic material with $\text{Re}\,\varepsilon_\parallel > 0$ and $\text{Re}\,\varepsilon_\perp < 0$. For small values of $f$ the metasurface is a dielectric.

The HTP resonance is most expressed in the hyperbolic phase of the metasurfaces, when the DBR has such a strong contrast that the coupling between fields on different sides of the DBR is excluded. In such a situation the spacer between the hyperbolic metasurface and the DBR serves as a metacavity supporting HTP. The optical properties of such a *hyperbolic metacavity* are demonstrated in Fig. 2, where the dependences of the reflection spectra in TM and TE polarizations on metal fraction $f$ are shown. The spectra feature a broadband stopgap for almost all plotted values of the parameters due to high contrast of the DBR ($n_h = 2.4$).

The HTP is visible as a strong resonance in the center of the stopgap in Figs. 2. Note that the reflectance in Figs. 2 (a)-(b) is in response to normal incidence with $\phi = 45°$. For a metasurface in the isotropic metal limit ($f = 1$) HTP reverts to a simple Tamm plasmon (TP) with an isotropic reflection minimum. The conversion of a TP into an HTP is accompanied by a considerable reduction in the resonance frequency. As metasurfaces are modified from an anisotropic metal to an anisotropic dielectric through the hyperbolic phase one can observe a dramatic drop in the HTP frequency of about 0.25 eV. This is related to the HTP resonance transition from the top of the stop gap through the lower-energy bound of the stop gap where the DBR becomes transparent. This coincides with the transition of the HTP into the regime of the Fabry-Perot (FP) resonance.

In Figs. 2 (a)-(b) we plot polarization ellipses on top of the HTP resonance, which represent the electric field in the middle of the metasurface layer. The positioning of the centers of the ellipses corresponds to the minima of the resonance reflectivity $|r_p|^2$, while the orientation of the ellipses is plotted with respect to the spatial *xy*-axes (with *xy*-directions indicated in Fig. 2). At the TP limit ($f = 1$) the metasurface is just a thin metal film, an isotropic material, and the electric field polarization follows the direction of excitation at $\phi = 45°$. But as $f$ is reduced, turning the metasurface into a hyperbolic material, the electric field inside of it is *pinned* to the *y*-direction perpendicular to the optic axis. In the hyperbolic phase the real part of the dielectric tensor component in the *y*-direction, $\mathrm{Re}\,\varepsilon_\perp < 0$, while along the optic axis $\mathrm{Re}\,\varepsilon_\parallel > 0$. Thus the pinning of the electric field in the hyperbolic phase corresponds to the metasurface polarizing along its metal-like axis.

The fact that the pinned electric field can be decomposed into TE and TM polarized fields (or $x'$- and $y'$-axes in the Fig. 2) shows that the response of the metasurface at the HTP resonance features a strong coupling between TM and TE polarizations. Both TM and TE polarized fields are featured in the HTP mode and it is described quite well by the equation of the metacavity resonance Eq. (5). This is shown by the blue curves in Fig. 2 (see also Fig. 1(a)). Those blue curves are found as minima of the right-hand side of Eq. (5). This demonstrates that HTPs serve as a good example of Type I metacavity resonances.

The TM/TE coupling provides for an additional channel to distribute the incoming power. As shown in our previous work if $f = 1$ the incident power is mostly *absorbed* in the TP mode due to the trapping of optical fields, despite of the small modal fraction within the metal[8]. Coupling to the TE field results in re-distribution of the power from absorption into TE reflection as can be seen in Fig. 2(b) where TE reflectivity is shown to be around 80% of the incidence power at the HTP resonance.

The appearance of TE polarized fields in response to TM excitation can be further understood from the perspective of the Polarization Rotator Equations (Eqs. (3)). For incidence planes along and perpendicular to the optic axis the response is purely TM polarized, but it is different due to the difference between $\varepsilon_\parallel$ and $\varepsilon_\perp$. According to Eqs.

(3) this leads to TE fields at intermediate angles $\phi$, with the biggest magnitude of $|r_s|^2$ and a minimum of $|r_p|^2$ near $\phi = \pi/4$, where HTP resonance is the strongest. To demonstrate this, we show the spectra of TM reflectivities $|r_p|^2$ and $\mathrm{Arg}\{r_p\}$ at $f = 0.52$ for $\phi = 0°, 45°, 90°$ in Figs. 3(a)-(c). At $\phi = 0°$ (Fig. 3(a)) the structure exhibits a strong reflectivity characteristic of the stop gap of the DBR, due to behavior of the metasurface as a dielectric with dielectric permittivity $\varepsilon_\parallel$ and $r_p(0°) \approx 1$. At $\phi = 90°$ (Fig. 3(c)) the metasurface acts as a metal in accordance with $\mathrm{Re}\,\varepsilon_\perp < 0$. The reflection coefficient exhibits a weak TP resonance in the middle to the stop gap with $r_p(90°) \approx -0.9$. This leads to the full cancellation of the TM polarized reflectivity at $\phi = 45°$, such that $r_p(45°) \approx 0$ at the HTP resonance according to Eqs. (3). Redistribution of the reflected power into TE polarized reflection is observed as was pointed out above (see Fig. 2(b)).

In the hyperbolic phase the HTP mode is highly dependent on the angle of the incidence plane $\phi$ and is only present within $\approx 30°$ FWHM angular ranges centered around $\phi = \frac{\pi}{4} + \frac{\pi m}{2}$, where $m$ is an integer. This is shown in Fig. 3(d) where the dispersion of HTP is shown. The presence of HTP can be judged by TM reflectivity being less than 50% within the green area and TE reflectivity being greater than 80% within the red-colored area. Thus HTPs feature a clover-leaf dispersion with highly anisotropic propagation properties. At other angles in the hyperbolic range the structure features a passive TM polarized 100% reflection.

We have previously demonstrated that TPs are related to Fabry-Perot (FP) resonances and continuously transform into them when a metal-DBR-metal (M-DBR-M) structure is modified into a metal-insulator-metal (MIM) structure via reduction of the dielectric contrast in the DBR[8]. It is interesting to see what happens to the HTPs when the same transformation is carried out. As can be seen from Fig 4(a), if the dielectric contrast in the DBR is reduced from $n_h/n_l = 3.6/2.4$ to $3.6/3.1$ there is a profound splitting of the resonance into two modes – lower-energy symmetric HTP (s-HTP) and anti-symmetric HTP (a-HTP). The magnetic field in the TM component of the symmetric (anti-symmetric) mode is predominantly symmetric (anti-symmetric) upon reflection with respect to the center of the structure. The splitting is due to hybridization of the HTPs at different metal-DBR interfaces.

**Type II Metacavities. Hyperbolic Fabry-Perot Resonances.** A further decrease of the dielectric contrast of the DBR causes the stop gap to collapse (see Fig. 4(b)) and there are no resonances of TP character. All of the resonances evolve into FP resonances as shown in Fig. 4(b). In the extreme metal ($f \approx 1$) and dielectric ($f \approx 0$) phases of the metasurfaces those resonances correspond to the usual FP modes without TM/TE coupling. Within the hyperbolic phase these resonances correspond to novel hyperbolic Fabry-Perot (HFP) modes whose existence we predict in this paper for the first time.

To understand the HFP resonances consider a Type II Metacavity, which is composed of a layer of homogeneous material sandwiched between two structures containing uniaxial metasurfaces with reflection coefficients $r_{pp}$ and $r_{sp}$ in TM polarization in response to TM and TE incidence correspondingly and $r_{ps}$ and $r_{ss}$ in TE polarization. In this case the resonance condition can be found from the following equation

$$\begin{pmatrix} A_p \\ A_p\, r_{pp}^L + A_s\, r_{sp}^L \\ A_s \\ A_p\, r_{ps}^L + A_s\, r_{ss}^L \end{pmatrix} = \begin{pmatrix} e^{i\varphi} & 0 & 0 & 0 \\ 0 & e^{-i\varphi} & 0 & 0 \\ 0 & 0 & e^{i\varphi} & 0 \\ 0 & 0 & 0 & e^{-i\varphi} \end{pmatrix} \begin{pmatrix} B_p\, r_{pp}^R + B_s\, r_{sp}^R \\ B_p \\ B_p\, r_{ps}^R + B_s\, r_{ss}^R \\ B_s \end{pmatrix}. \qquad (6)$$

Eq. (6) is structured similarly to Eq. (4) and has nontrivial solutions only if

$$\left(r_{pp}^L r_{pp}^R + r_{sp}^L r_{ps}^R + \frac{(r_{pp}^L r_{sp}^R + r_{sp}^L r_{ss}^R)(r_{pp}^R r_{ps}^L + r_{ps}^R r_{ss}^L)e^{2i\varphi}}{1 - r_{ps}^L r_{sp}^R e^{2i\varphi} - r_{ss}^L r_{ss}^R e^{2i\varphi}}\right) e^{2i\varphi} - 1 = 0,$$

$$\left(r_{ss}^L r_{ss}^R + r_{ps}^L r_{sp}^R + \frac{(r_{pp}^L r_{sp}^R + r_{sp}^L r_{ss}^R)(r_{pp}^R r_{ps}^L + r_{ps}^R r_{ss}^L)e^{2i\varphi}}{1 - r_{pp}^L r_{pp}^R e^{2i\varphi} - r_{sp}^L r_{ps}^R e^{2i\varphi}}\right) e^{2i\varphi} - 1 = 0.$$

(7)

As can be seen from Eq. (7) the resonances of the metacavities involve co-propagation and multiple coupled reflections of TM and TE polarized waves in a similar, but a more complex form, than the one shown in Fig 1(a). We plot the minima of the right-hand side of Eq. (7) in Fig. 4(b) as set of blue dots and they correspond nicely to the HFP resonance minima in reflectivity $|r_p|^2$ of the structure.

The fields in HFP resonances exhibit unique properties. In Fig. 5(a) we show the electric field in the metasurface cavity with parameters indicated by a black dot in Fig. 4(b). The field inside of the cavity is a standing wave, with polarization rotating between the sides of the cavity, but being close to linear at any given point inside of the cavity. This is demonstrated by the blue curve in Fig. 5(a), which follows the end of the electric field vector through the metacavity at a given moment of time, meanwhile in red the polarization ellipsoids are shown at several locations and demonstrate the close to linear polarization at those points. One can notice an efficient 90° polarization rotation as light passes through this Type II Metacavity.

In Fig. 5(b) we summarize the transition from the HTP resonance to HFP resonance as the contrast in the DBR is reduced (i.e. $n_l$ is increased to $n_h = 3.6$). The pie charts in the figure illustrate the power channels over which the incoming power is distributed for the values of the parameters at which they are shown, with "*rp*", "*rs*", "*tp*" and "*ts*" being the TM and TE reflections and TM and TE transmissions correspondingly. Absorption is shown as the blue contribution, but never marked by a letter since it is small.

At high contrast (shown in the bottom of Fig. 5(b)) the cavity has a broad stopgap and HTP resonance can be seen at $\hbar\omega = 1.18$ eV. As was described above about 80% of the incoming TM polarized power goes into TE reflection, indicating that HTP-supporting structures can serve as efficient $90°$ polarization rotation mirrors. As the contrast is reduced the HTP splits and within the resulting s-HTP and a-HTP resonances the power is distributed over all four channels (with exception of absorption which is small). At smaller contrasts the stop gap is not as broad and the split modes leave the stop gap becoming HFP modes in nature. Two other HFP modes are shown on the sides of Fig. 5(b). In these modes power also is distributed over all four reflection and transmission channels. They converge and cross with the previously mentioned HFPs and interfere with them. This results in the predominance of TE transmission as indicated at the intersection of the modes (top of Fig. 5(b)).

In conclusion, we propose a new class of optical resonators - *metacavities* - and predict that such resonators support novel *hyperbolic resonances.* These hyperbolic optical modes feature unique electromagnetic field distribution within the metacavity, strong TM/TE polarizations coupling and anisotropic dispersion, offering new avenues of photonics research.

**Acknowledgments**

This study was supported by funds from the Office of the Vice President for Research & Economic Development and the Jack N. Averitt College of Graduate Studies at Georgia Southern University. D. K. is grateful for the support provided by the student research grant from College Office of Undergraduate Research (COUR) at Georgia Southern University.

**Materials & Correspondence**. Correspondence and requests for materials should be addressed to M. D. over email: *mdurach@georgiasouthern.edu*.

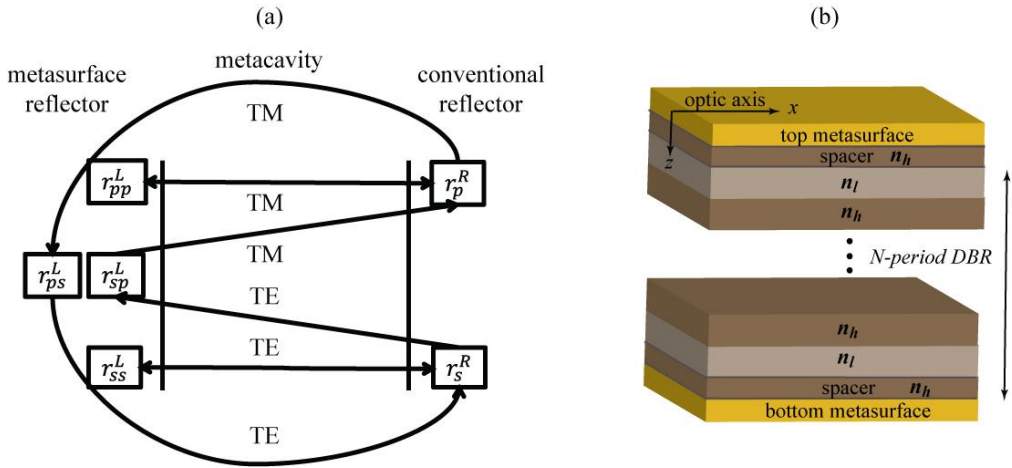

**Fig. 1. Resonances of the metacavities. (a)** Schematic of the resonances in Type I Metacavities **(b)** Schematic of the example structure, which is a Type I Metacavity at strong contrast in the DBR and a Type II Metacavity at no contrast.

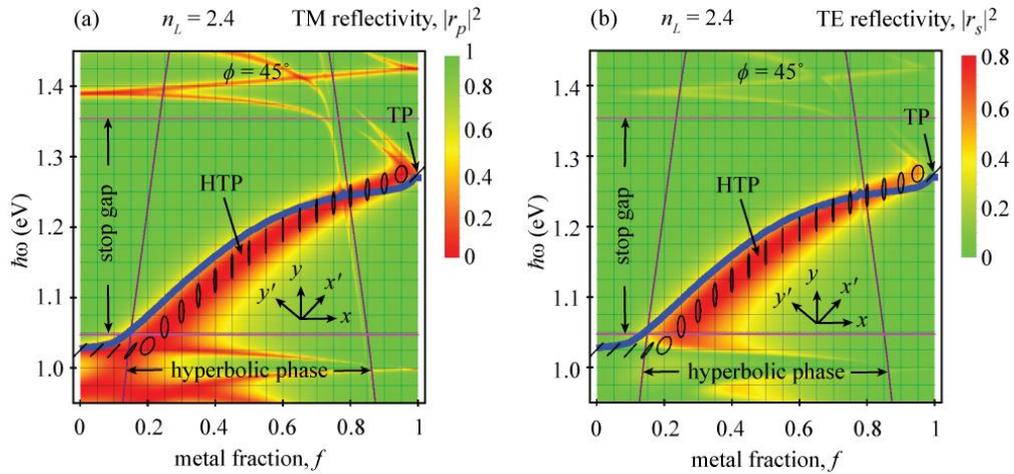

**Fig. 2. Optical properties of the structure supporting HTPs. (a)** TM reflectivity, $|r_p|^2$, as a function of excitation frequency $\hbar\omega$ and metal fraction $f$; **(b)** the same as (a), but for TE reflectivity, $|r_s|^2$.

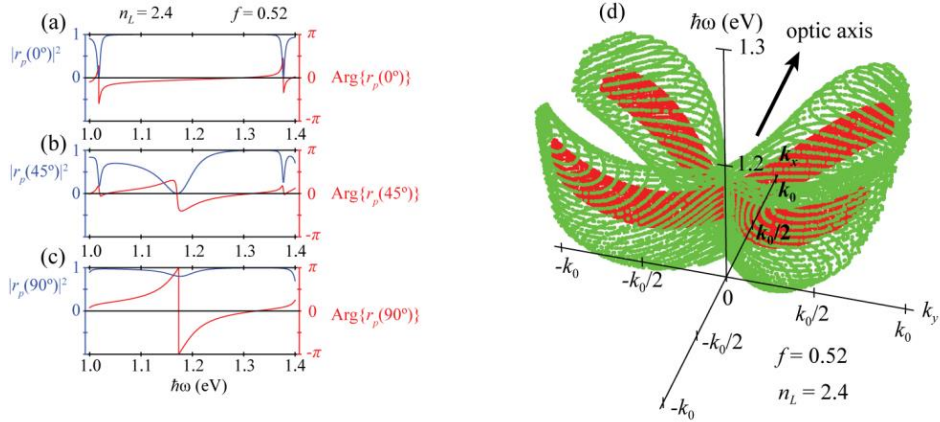

**Fig. 3. Anisotropic nature of HTPs. (a)-(c)** Formation of HTP resonance at $\phi = 45°$ due to difference of reflectivity at $\phi = 0°$ and $\phi = 90°$ in accordance with Polarization Rotation Equations (Eqs. (3)). **(d)** Clover-leaf parabolic dispersion of HTP modes.

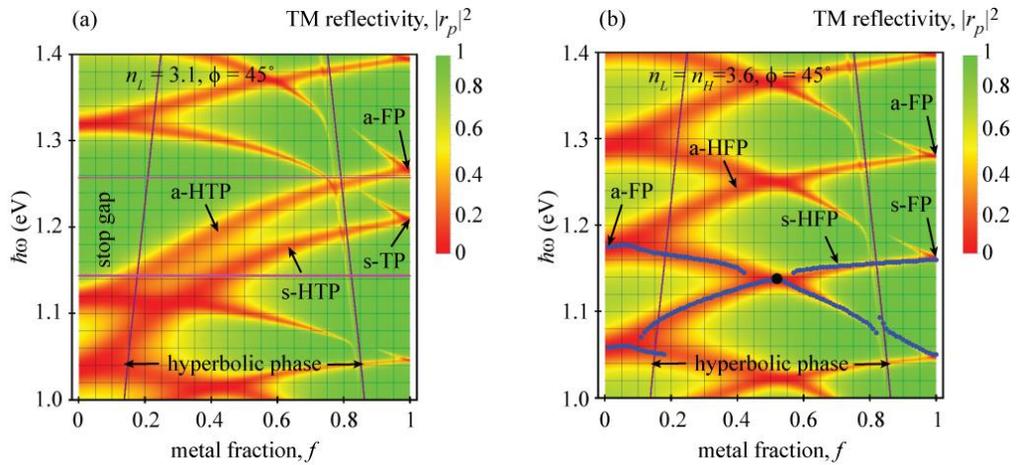

**Fig. 4. Splitting of HTPs and formation of HFP resonances. (a)** TM reflectivity spectrum, $|r_p|^2$, at intermediate contrast in the DBR ($n_l = 3.1$), **(b)** The same at no contrast in the DBR.

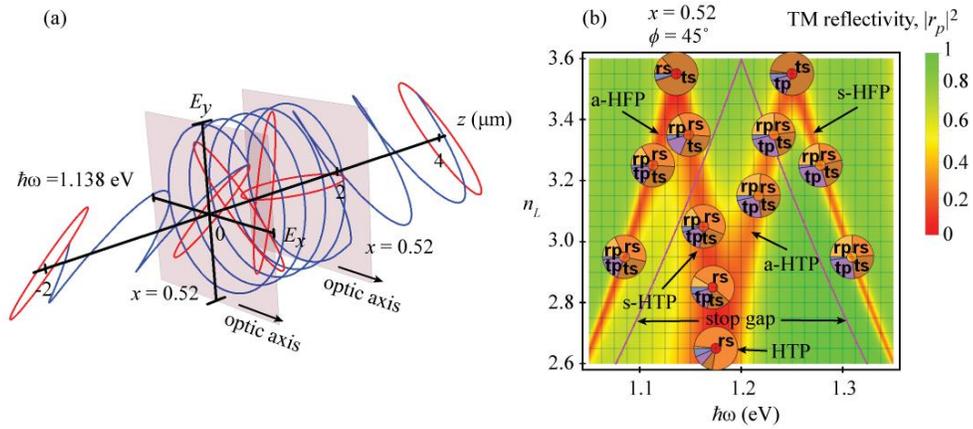

**Fig. 5. Electric field and power distribution channels in hyperbolic resonances. (a)** Electric field at HFP in a Type II Metacavity. **(b)** Evolution of power distribution in Hyperbolic Metacavities.

**Supplementary Information**

**Characteristic Matrix Approach.** Consider a uniaxial anisotropic medium whose optical axis is in xy-plane at angle ϕ with respect to the x-axis. The dielectric permittivity tensor is

$$\hat{\epsilon} = \begin{pmatrix} \epsilon_\parallel \cos\phi^2 + \epsilon_\perp \sin\phi^2 & (\epsilon_\perp - \epsilon_\parallel)\cos\phi\sin\phi & 0 \\ (\epsilon_\perp - \epsilon_\parallel)\cos\phi\sin\phi & \epsilon_\perp \cos\phi^2 + \epsilon_\parallel \sin\phi^2 & 0 \\ 0 & 0 & \epsilon_\perp \end{pmatrix} = \begin{pmatrix} \epsilon_{11} & \epsilon_{12} & 0 \\ \epsilon_{12} & \epsilon_{22} & 0 \\ 0 & 0 & \epsilon_\perp \end{pmatrix} \quad (8)$$

The Maxwell equations can be written as $E_z = -XH_y/\epsilon_\perp$, and $H_z = XE_y$, where $X = k/k_0$. Also $E'_x = ik_0 H_y - ik^2 H_y/(k_0 \epsilon_\perp)$, $H'_y = ik_0 \epsilon_{11} E_x + ik_0 \epsilon_{12} E_y$, $E'_y = -ik_0 H_x$, $H'_x = -ik_0 \epsilon_{12} E_x + ik^2 E_y/k_0 - ik_0 \epsilon_{22} E_y$. This can be collected as

$$\frac{\partial}{\partial z}\psi = ik_0 \hat{\Delta}\, \psi, \quad (9)$$

where $\boldsymbol{\psi} = (H_y, E_x, E_y, H_x)$ and

$$\hat{\Delta} = \begin{pmatrix} 0 & \epsilon_{11} & \epsilon_{12} & 0 \\ 1 - X^2/\epsilon_\perp & 0 & 0 & 0 \\ 0 & 0 & 0 & -1 \\ 0 & -\epsilon_{12} & -(\epsilon_{22} - X^2) & 0 \end{pmatrix}. \quad (10)$$

The solution has a form $\psi = \psi_0 \exp(iqz)$, which leads to a characteristic equation

$$\hat{\Delta}\psi = Q\psi, \quad (11)$$

were $Q = q/k_0$. The eigenvalues are $Q = \pm Q_o = \pm\sqrt{\epsilon_\perp - X^2}$ and $Q = \pm Q_e = \pm\sqrt{\epsilon_\parallel - X^2\left(\sin\phi^2 + \frac{\epsilon_\parallel}{\epsilon_\perp}\cos\phi^2\right)}$.

From this one can obtain a characteristic matrix $\widehat{M}$ of such an anisotropic layer in the region $0 < z < d$

$$\widehat{M} = \frac{1}{1 + Y(\tan\phi)^2} \left(\widehat{M}_0 + \left(\widehat{M}_{1e} - \widehat{M}_{10}\right)\tan\phi + \widehat{M}_2(\tan\phi)^2\right), \tag{12}$$

such that

$$\begin{pmatrix} E_x(z=0) \\ H_y(z=0) \\ E_y(z=0) \\ -H_x(z=0) \end{pmatrix} = \widehat{M} \cdot \begin{pmatrix} E_x(z=d) \\ H_y(z=d) \\ E_y(z=d) \\ -H_x(z=d) \end{pmatrix}.$$

The matrices in Eq. (12)

$$\widehat{M}_0 = \begin{pmatrix} \cos(Q_e k_0 d) & -iYQ_e \sin(Q_e k_0 d) & 0 & 0 \\ -\frac{i}{YQ_e}\sin(Q_e k_0 d) & \cos(Q_e k_0 d) & 0 & 0 \\ 0 & 0 & \cos(Q_0 k_0 d) & \frac{i}{Q_0}\sin(Q_0 k_0 d) \\ 0 & 0 & iQ_0 \sin(Q_0 k_0 d) & \cos(Q_0 k_0 d) \end{pmatrix},$$

$$\widehat{M}_{1e} = \begin{pmatrix} 0 & 0 & iYQ_e \sin(Q_e k_0 d) & Y\cos(Q_e k_0 d) \\ 0 & 0 & -\cos(Q_e k_0 d) & -\frac{i}{Q_e}\sin(Q_e k_0 d) \\ \frac{i}{Q_e}\sin(Q_e k_0 d) & -Y\cos(Q_e k_0 d) & 0 & 0 \\ \cos(Q_e k_0 d) & -iYQ_e \sin(Q_e k_0 d) & 0 & 0 \end{pmatrix},$$

$$\widehat{M}_{10} = \begin{pmatrix} 0 & 0 & iYQ_0 \sin(Q_0 k_0 d) & Y\cos(Q_0 k_0 d) \\ 0 & 0 & -\cos(Q_0 k_0 d) & -\frac{i}{Q_0}\sin(Q_0 k_0 d) \\ \frac{i}{Q_0}\sin(Q_0 k_0 d) & -Y\cos(Q_0 k_0 d) & 0 & 0 \\ \cos(Q_0 k_0 d) & -iYQ_0 \sin(Q_0 k_0 d) & 0 & 0 \end{pmatrix},$$

$$\widehat{M}_2 = \begin{pmatrix} Y\cos(Q_0 k_0 d) & -iY^2 Q_0 \sin(Q_0 k_0 d) & 0 & 0 \\ -\dfrac{i}{Q_0}\sin(Q_0 k_0 d) & Y\cos(Q_0 k_0 d) & 0 & 0 \\ 0 & 0 & Y\cos(Q_e k_0 d) & \dfrac{iY}{Q_e}\sin(Q_e k_0 d) \\ 0 & 0 & iYQ_e \sin(Q_e k_0 d) & Y\cos(Q_e k_0 d) \end{pmatrix},$$

where $Y = \dfrac{\epsilon_\perp}{\epsilon_\perp - X^2}$.

**Polarization Rotation Equations.** Consider a coordinate system in which $x'$-axis is directed along the incidence plane, while $x$-axis is along the optical axis. The incident wave in different coordinates are related as

$$\begin{pmatrix} E^{inc}_{x(0°)} \\ E^{inc}_{x(90°)} \end{pmatrix} = \widehat{M} \begin{pmatrix} E^{inc}_{x'(TM)} \\ E^{inc}_{y'(TE)} \end{pmatrix} = \widehat{M}\begin{pmatrix} p_{p0} \\ 0 \end{pmatrix}, \quad \widehat{M} = \begin{pmatrix} \cos\phi & -\sin\phi \\ \sin\phi & \cos\phi \end{pmatrix}$$

At the same time the reflected and transmitted waves can be found as

$$\begin{pmatrix} E^{ref}_{x(0°)} \\ E^{ref}_{x(90°)} \end{pmatrix} = \begin{pmatrix} -r_p^{0°} & 0 \\ 0 & -r_p^{90°} \end{pmatrix}\begin{pmatrix} E^{inc}_{x(0°)} \\ E^{inc}_{x(90°)} \end{pmatrix}, \quad \begin{pmatrix} E^{tr}_{x(0°)} \\ E^{tr}_{x(90°)} \end{pmatrix} = \begin{pmatrix} p_{ps}t_p^{0°}/p_{p0} & 0 \\ 0 & p_{ps}t_p^{90°}/p_{p0} \end{pmatrix}\begin{pmatrix} E^{inc}_{x(0°)} \\ E^{inc}_{x(90°)} \end{pmatrix}$$

Combining these relationships

$$\begin{pmatrix} E^{ref}_{x'(TM)} \\ E^{ref}_{y'(TE)} \end{pmatrix} = \begin{pmatrix} -p_{p0}r_p^\phi \\ r_s^\phi \end{pmatrix} = \widehat{M}^{-1}\begin{pmatrix} E^{ref}_{x(0°)} \\ E^{ref}_{x(90°)} \end{pmatrix} = \widehat{M}^{-1}\begin{pmatrix} -r_p^{0°} & 0 \\ 0 & -r_p^{90°} \end{pmatrix}\widehat{M}\begin{pmatrix} p_{p0} \\ 0 \end{pmatrix}$$

$$\begin{pmatrix} E^{tr}_{x'(TM)} \\ E^{tr}_{y'(TE)} \end{pmatrix} = \begin{pmatrix} p_{ps}t_p^\phi \\ t_s^\phi \end{pmatrix} = \widehat{M}^{-1}\begin{pmatrix} E^{tr}_{x(0°)} \\ E^{tr}_{x(90°)} \end{pmatrix} = \widehat{M}^{-1}\begin{pmatrix} p_{ps}t_p^{0°}/p_{p0} & 0 \\ 0 & p_{ps}t_p^{90°}/p_{p0} \end{pmatrix}\widehat{M}\begin{pmatrix} p_{p0} \\ 0 \end{pmatrix}$$

From which using an identity

$$\widehat{M}^{-1}\begin{pmatrix}a_1 & 0 \\ 0 & a_2\end{pmatrix}\widehat{M} = \begin{pmatrix}a_1\cos^2\phi + a_2\sin^2\phi & (a_2-a_1)\sin\phi\cos\phi \\ (a_2-a_1)\sin\phi\cos\phi & a_1\sin^2\phi + a_2\sin^2\phi\end{pmatrix}$$

we obtain Eqns. (3).